\documentclass[12pt,a4paper,twoside]{article}
\usepackage{amssymb}
\usepackage{epsfig}

\usepackage{graphicx}
\usepackage{amsmath}
\usepackage[widemargins]{a4}

\def\dint{\mathop{\displaystyle \int}}%

\begin{document}

\title{Looking for new  solutions to the hierarchy problem}
\author{Francesco Caravaglios \\
Dipartimento di Fisica, Universit\`a di Milano, \textit{and } INFN sezione
di Milano}
\date{May 20, 2002}
\maketitle

\begin{abstract}
While in first and second 
quantization the fundamental operators   are respectively coordinates and 
fields (functions), an extension of quantum field theory can be achieved if 
the usual   pair of  conjugate momenta is
 represented  by functionals.
After a brief introduction on the hierarchy problem,  we show  how 
the ordinary
 quantum field theory  can arise from a specific limit of this  extension.
We will also show how  this extension can offer new solutions to the 
hierarchy problem. 
The peculiarity that makes this scenario appealing (as possible solution of
 the hierarchy problem), is the  absence of new 
 light particles at (or close to) the electroweak scale. This is in much 
better   agreement  with the experimental observation, since (until now)  
all searches for  new physics signals have confirmed the remarkable success of
 the Standard Model (contrary to common expectations).

\end{abstract}

\section{Introduction}

\noindent The incredible experimental success of the renormalizable \cite{ren}
 Standard
Model \cite{sm}, well beyond any theoretical expectation, puts 
severe constraints to
any attempt to solve  the hierarchy problem introducing new particles or
states at the weak scale. All experiments confirm  the Standard Model. It is
true that all these experiments could have missed new light particles
signals for some fortuitous coincidence or for some theoretical and still
unknown reason,  but we should not neglect the impressive convergence of
all experiments towards the same conclusion: the spectrum and the
properties of the low energy theory coincide with that of the renormalizable
Standard Model. We stress the word renormalizable, because  it is ruled out
not only the real production of such new light particles but also their virtual
exchange, that could appear under the form of new effective non
renormalizable operators. Limits to these operators can be found in the
literature. They arise from precision electroweak data, from FCNC tests,
but also from proton stability and neutrino masses. Here we argue that this
experimental scenario  discloses a clue that  can help us to guess how to
extend the Standard Model.

\subsection{The problem}

With the (renormalizable) Standard Model Hamiltonian, we are able to
compute the scattering matrix for processes that involve different type of  
$|in>$ and $|out>$ states.  All experimental tests prove the amazing
accuracy of such predictions. In particular, experiments rule out the
possibility to modify the renormalizable Standard Model Hamiltonian either
adding new light particles or introducing new higher dimensional operators
with a characteristic scale $\Lambda $ (according to the operator choice, $%
\Lambda $ can range from the TeV scale, up to the Grand Unification Scale).
On the other hand the same Hamiltonian is not able to give a realistic
estimate of the effective potential, explaining the origin of the Hierarchy
between the weak scale and the unification scale. The mass parameter in the
Higgs potential is unstable under radiative corrections and it depends
quadratically on the cut-off $\Lambda $, the scale of the new physics.
Theories that introduces this cut-off close to the weak scale directly to
the Hamiltonian, do not explain why present experiments have not yet found
evidence for departures from the Standard Model. Theorists acknowledge  the
severe constraints to a relatively light new physics scale,  and cope with
them,  assuming  new symmetries that could reduce the impact of such \
light scale to the   experimental tests, and getting used to the idea that
a certain amount of fine tuning  must be accepted.  This scenario becomes
less puzzling, if we look closer at the physical quantities that we are
comparing. From one side we have very accurate predictions of the scattering
matrix elements, between $|in>$ and $|out>$ states with\footnote{$\phi $ is
the Higgs field.} the average $<in,out|\,\phi \,|in,out>$  very close to $v$%
, the  electroweak vev . These matrix elements, and the time evolution of
these states are very well described by the Standard Model. On the other
hand, we know that the same Standard Model hamiltonian is unable to evaluate
the energy of states with a very different vev $<\psi |\phi |\psi >=V>>v$ (
i.e. the SM does not give an acceptable effective potential). However this
set of states is experimentally very different from the previous one, since
a state with $<\psi |\phi |\psi >=V$ cannot  be reproduced in a
laboratory. To our opinion these considerations raise a doubt: can we
extrapolate ordinary quantum field theory to include the time evolution of
these states, with very different  vev, and very far from our experience?
In the following we will try to see if we can give a theoretical basis to
this question.

\subsection{A theoretical  clue to solve the Hierarchy problem: the
mathematical example of the non linear Schr\"{o}dinger equation}

It is useful to exemplify the problem as follows: we show that a non linear
Schr\"{o}dinger equation could explain an apparent paradox in a similar
situation but in the simpler context  of first
quantization. To be concrete we take an explicit mathematical example.
Consider the following equation 
\begin{equation}
i\frac{\partial }{\partial t}\,\,\psi (x,t)=-\frac{\partial ^{2}}{\partial
x^{2}}\,\psi (x,t)+x^{2}\,\ \ V\left( \int \psi ^{\ast
}(y,t)\,\,y\,^{2}\,\psi (y,t)\,dy\right) \ \psi (x,t).  \label{nonlinear}
\end{equation}
Where $V(x)$ is an analytic real function with $V>0$, and $V(x)$ going
exponentially to zero when $x\rightarrow \infty .$ The solutions of this
equation  conserve the probability, and for any $t>0$ we have (if \ $\psi $
is normalized, at $t=0$) 
\begin{equation}
\int \psi ^{\ast }(x,t)\,\,\,\psi (x,t)\,dx=1
\end{equation}
$\ $This is simple to check once we observe that $\int \psi ^{\ast
}(y,t)\,\,y^{2}\,\psi (y,t)\,dy$, and its time derivatives,  are real for
any $t$. Then 
\begin{equation}
\frac{\partial }{\partial t}\int \psi ^{\ast }(x,t)\,\,\,\psi (x,t)\,dx=\int
\psi ^{\ast }(x,t)\,\,\frac{\partial }{\partial t}\,\psi (x,t)\,dx+\int \psi
(x,t)\frac{\partial }{\partial t}\,\,\,\psi ^{\ast }(x,t)\,dx=0
\end{equation}
where we have used (\ref{nonlinear}) . One can also verify that the energy 
\begin{equation}
E=-\int \psi ^{\ast }(x,t)\frac{\partial ^{2}}{\partial x^{2}}\,\psi (x,t)\,\
dx+F\left( \int \psi ^{\ast }(y,t)\,\,y\,^{2}\,\psi (y,t)\,dy\right) 
\end{equation}
with $F^{\prime }=V$ is conserved. One can study the class of solutions of
this equation that satisfy $\int \psi ^{\ast }(y,t)\,\,y^{2}\,\psi
(y,t)\,dy<<V(0)/V^{\prime }(0).$ In this approximation the (\ref{nonlinear})
is an ordinary Schr\"{o}dinger Equation  with the potential of an harmonic
oscillator (and with bounded states) provided that we replace $%
V(x)\rightarrow V(0).$ On the other hand, since $V(x)$  exponentially goes
to zero when $x\rightarrow \infty $,  we can also have solutions that are
similar to a free gaussian wave packet, centered in $x>>1$ and going to
infinity when $t$ goes to infinity.  This situation is similar to the
puzzle that we  have  discussed in the previous section, where the
experimental scenario apparently required two distinct Hamiltonians with two
different class of solutions: one with small vev and another one with very
large vev. Clearly a unique linear equation cannot describe the evolution of
the physical system in both physical regions, and only a non linear equation
can correctly describe the time evolution in all regions. In the following
we would like to see if this \textit{naive }idea can be  exploited to solve
the hierarchy problem (taking into account all necessary modifications
required by the different physical context).  

\section{Second Quantization: Definitions}

To start, we remind the formalism and the basics of the second quantization
of a scalar field. The classical action  for a free scalar field is 
\begin{equation}
\mathcal{A}=\dint \frac{1}{2}(-\phi \,\partial ^{2}\phi -m^{2}\phi
^{2})\,d^{4}x
\end{equation}
 from which we get the Hamiltonian 
\begin{equation}
\mathcal{H}=\dint \frac{1}{2}(\pi ^{2}(x)-\phi (x)\,\triangledown ^{2}\phi
(x)+m^{2}\phi ^{2}(x))\,d^{3}x.
\label{hami1}
\end{equation}
The rule of second quantization  imposes replacing $\phi $ and $\pi $ with
field operators $\hat{\phi}$ and $\hat{\pi}$ satisfying the equal time
commutation relations below 
\begin{equation}
\lbrack \,\hat{\pi}(x)\,,\,\hat{\phi}(y)\,]=[\,\,\dot{\phi}(y),\,\hat{\phi}%
(x)\,]=i\text{%
h\hskip-.2em\llap{\protect\rule[1.1ex]{.325em}{.1ex}}\hskip.2em%
}\delta ^{3}(x-y).  \label{commutation}
\end{equation}
\bigskip We can give an explicit representation of the commutation 
(\ref{commutation}) in the space of functionals\footnote{%
For convention, $S[\phi ]$ (with square brackets) denotes a functional of
the real function $\phi (x)$, while $S(\phi )$   denotes an ordinary
function $S$ of the real number $\phi $.} $S[\phi ]$ with $\phi $ being a \
real function $\phi (x)$ of the three dimensional space. The product of two
states $<\psi _{1}|\psi _{2}>$  can be (formally) defined in terms of a
path integral (i.e. a functional integration) as follows\footnote{%
Note the analogy with first quantization where the  product of  $<\psi
_{1}|\psi _{2}>=\int \psi _{1}^{\ast }(x)\,\ \psi _{2}(x)\,dx$. The
functional integration is defined as usual \cite{zuber}; an infinite
constant factor is always understood to get a well defined and finite
result. We will come back to this infinite constant after.} 
\begin{equation}
<\psi _{1}|\psi _{2}>=\dint \mathcal{D}\phi S_{1}^{\dagger }[\phi
]S_{2}[\phi ].  \label{product}
\end{equation}
With these definitions, it is easy to guess the action of $\ \,\hat{\pi}(x)\,
$ and $\ \hat{\phi}(y)$ onto a state $|S>$. Namely 
\begin{equation}
\hat{\phi}(y)\,\ |S>\Rightarrow \phi (y)\,S[\phi ]
\label{replace0}
\end{equation}
and 
\begin{equation}
\hat{\pi}(y)\,\ |\psi >\Rightarrow i\frac{\delta }{\delta \phi (y)}\,S[\phi ]
\label{replace}
\end{equation}
where $i$%
h\hskip-.2em\llap{\protect\rule[1.1ex]{.325em}{.1ex}}\hskip.2em%
$\delta /\delta \phi (y)$ is a functional derivative. One can check that 
\begin{equation}
\lbrack i\text{%
h\hskip-.2em\llap{\protect\rule[1.1ex]{.325em}{.1ex}}\hskip.2em%
}\,\frac{\delta }{\delta \phi (y)},\phi (x)]=i\text{%
h\hskip-.2em\llap{\protect\rule[1.1ex]{.325em}{.1ex}}\hskip.2em%
}\delta ^{3}(x-y)  \label{comm1}
\end{equation}
 as required by the (\ref{commutation}). With the replacement
 (\ref{replace0},\ref{replace}),
the Hamiltonian (\ref{hami1}) in this functional representation becomes 
\begin{equation}
H=\dint \frac{1}{2}(-\text{%
h\hskip-.2em\llap{\protect\rule[1.1ex]{.325em}{.1ex}}\hskip.2em%
}^{2}\frac{\delta ^{2}}{\delta \phi (x)^{2}}-\phi \,\triangledown ^{2}\phi
+m^{2}\phi ^{2})d^{3}x.  \label{hamifunc}
\end{equation}
This functional operator leads to the Schr\"{o}dinger functional equation 
\begin{equation}
\dint \frac{1}{2}(-\text{%
h\hskip-.2em\llap{\protect\rule[1.1ex]{.325em}{.1ex}}\hskip.2em%
}^{2}\frac{\delta ^{2}S[\phi ]}{\delta \phi (x)^{2}}-(\phi \,\triangledown
^{2}\phi )S[\phi ]+m^{2}\phi ^{2}S[\phi ])d^{3}x=E\,S[\phi ]
\label{eq:scroe}
\end{equation}
 where $S$ is a generic eigenfunctional of $H$ with eigenvalue $E$. The
state with minimal $E$ can be written\footnote{%
The eigenvalue $E$ is divergent, but this divergence can be removed by a
proper subtraction procedure.} 
\begin{equation}
S_{0}[\phi ]=N\exp (-\frac{1}{2}\int d^{3}x\,\phi (x)\sqrt{-\triangledown
^{2}+m^{2}}\phi (x))  \label{vacuum}
\end{equation}
 with a (divergent) normalization constant $N$ that sets

\begin{equation}
\dint \mathcal{D}\phi S_{0}^{\dagger }[\phi ]S_{0}[\phi ]=1.
\label{normalizzazione}
\end{equation}
The above integration $\mathcal{D}\phi $ is understood in the functional
sense \cite{zuber} as in common path integrals. One can recognizes that $S_{0}[\phi ]$,
as given in (\ref{vacuum}), is similar to the wave function of the common
harmonic oscillator, apart from the obvious change of functions with
functionals. Starting from the vacuum $S_{0}$we can build other
states , for instance we can create a new state through the action of the
operator $\hat{\phi}$  at the position $z$, $\hat{\phi}(z)\,\ |0>$ and the
corresponding functional becomes 
\begin{equation}
S_{z}[\phi ]=N\,\ \phi (z)\exp (-\int d^{3}x\,\phi (x)\sqrt{-\triangledown
^{2}+m^{2}}\phi (x))  \label{oneparticle}
\end{equation}

We can also compute the product 
\begin{equation}
<0|\phi ^{\dagger }(z)\,\phi (y)\,|0>=\dint \mathcal{D}\phi S_{z}^{\dagger
}[\phi ]S_{y}[\phi ]=N^{2}\dint \mathcal{D}\phi \,\phi (x)\phi (y)\exp
(-\int d^{3}x\,\phi (x)\sqrt{-\triangledown ^{2}+m^{2}}\phi (x))
\end{equation}

This is easily evaluated once we realize that this is simply the definition
of the two point Green function of three dimensional (euclidean) scalar
theory with inverse propagator $\sqrt{-\triangledown ^{2}+m^{2}}.$ This
yields 
\begin{equation}
\frac{1}{2\sqrt{-\triangledown _{z}^{2}+m^{2}}}\delta ^{3}(z-y)=\int \frac{%
d^{3}p}{2(2\pi )^{3}}\frac{e^{ip\cdot (z-y)}}{\sqrt{p^{2}+m^{2}}}
\end{equation}
      and this is the same result that one gets taking  the scalar
propagator at equal times.  It is important to note that the constant $N$
used in (\ref{vacuum}) and in (\ref{oneparticle}) is the same: thus the product 
(\ref{product}) is well defined once we have chosen a vacuum functional (\ref
{vacuum}), that  sets the normalization $N$  (\ref{normalizzazione}), once for
all.

We stress that  until now we have not introduced any new physics or new
physical concept. We have only introduced a rather unusual formalism  to
define second quantization. Our aim, here, is to put in better evidence the
parallel  between the well known Schr\"{o}dinger equation in first
quantization and its (less popular) analogue in second quantization (\ref
{eq:scroe}). This will become useful after, because it will make more
transparent   some essential clues of the hierarchy problem that will lead
us to naturally  extend quantum field theory. \

\section{\protect\bigskip Quantization of  free functional fields}

In what follows we will only introduce some  basic concepts concerning the 
 quantization of functional fields. We do not have the intention to be
comprehensive, and sometimes we will be not rigorous. Even if the way to
proceed is not unique, some important conclusions concerning the hierarchy
problem can be achieved without entering into difficult theoretical details.
In the discussion below, we will often put in evidence the parallel with
first and second quantization,  this  will help us to guess how some
results of these well known theories could be generalized into some sort of 
 third quantization.
Namely a theory where the role of fields $\psi(x)$ is now played by functionals
 $S[\phi ]$.
\bigskip Note that eq. (\ref{eq:scroe}) can come from the action $\mathcal{A}$
below, where the integral $\mathcal{D}\phi $ is taken in the functional
sense 
\begin{equation}
\mathcal{A=}\dint \mathcal{D}\phi \,S^{\dagger }[\phi ,t]\left( i\frac{%
\partial }{\partial t}+\frac{1}{2}\left( -\text{%
h\hskip-.2em\llap{\protect\rule[1.1ex]{.325em}{.1ex}}\hskip.2em%
}^{2}\frac{\delta ^{2}}{\delta \phi (x)^{2}}-(\phi \,\triangledown ^{2}\phi
)+m^{2}\phi ^{2}\right) \right) S[\phi ,t]\,d^{3}x\,dt+\text{h.c.}
\label{fund}
\end{equation}
In fact, if we formally define a functional derivative , such that 
\begin{equation}
\frac{\delta }{\delta S[\phi ^{\prime }]}S[\phi ]=\prod_{X}\delta \left(
\phi (X)-\phi ^{\prime }(X)\right) \equiv \delta ^{\infty }[\phi -\phi
^{\prime }]  \label{delta1}
\end{equation}
and 
\begin{equation}
\frac{\delta }{\delta S[\phi ^{\prime }]}\int \mathcal{D}\phi \,S[\phi
]G[\phi ]=G[\phi ^{\prime }],  \label{delta}
\end{equation}
the equation $\delta \mathcal{A}/\delta \,S^{\dagger}[\phi ]=0$ 
is equivalent to the (\ref
{eq:scroe}). Let us try to quantize the action $\mathcal{A}$. As usual the
momentum conjugate of $S[\phi ]$ is obtained taking the derivative of $%
\mathcal{A}$ with respect $\dot{S}[\phi ]$. Thus $S[\phi ]$  and $%
i\,S^{\dagger }[\phi ]$  represent a couple of conjugate momenta, that for
the rule of  quantization must obey the equal time anti-commutation
prescription (if we consider fermion-like functional operators)

\begin{equation}
\{iS^{\dagger }[\phi ,t],S[\phi ^{\prime },t]\}=\prod_{i=1,N}\delta
^{k}\left( \phi (X_{i})-\phi ^{\prime }(X_{i})\right) \equiv \delta ^{\infty
}[\phi -\phi ^{\prime }]  \label{anticom}
\end{equation}
As expected, the space-time Dirac delta is replaced by a functional Dirac
delta. The Hamiltonian for these \textit{free} \textit{functional  }fields 
$S$
is given by 
\begin{equation}
\mathcal{H=}\dint \mathcal{D}\phi \,S^{\dagger }[\phi ]\,H\,S[\phi ]
\label{hami}
\end{equation}
where $H$ is a functional operator given by (\ref{hamifunc}), in the representation (\ref{replace0},\ref{replace}).

We can also define the operator 
\begin{equation}
\mathcal{N=}\dint \mathcal{D}\phi \,S^{\dagger }[\phi ]\,\ S[\phi ]
\end{equation}
that commutes with the hamiltonian $\mathcal{H}$. From the commutation
relation $\left[ \mathcal{N\,},S[\phi ]\right] =S[\phi ]$, we recognize that
\ $S^{\mathbf{\dagger }}$ and $S$ are creation and annihilation \textit{%
functional } operators. The eigenstates of the  hamiltonian (\ref{hami}) \
can be found, following the same arguments used in second quantization, \
and exploiting the commutativity of  the Hamiltonian with the number of
fields operator .$\mathcal{N}$ . Namely, we start from the vacuum $|0>$,
defined by 
\begin{equation}
S|0>=0
\end{equation}
 from which we get 
\begin{equation}
\mathcal{H}|0>=0
\end{equation}
The vacuum is an eigenstate of the Hamiltonian with zero energy. Then we
proceed building all the other states. We can apply iteratively and several
times the creation \textit{functional } operator. For example, the one 
\textit{functional} creation operator $S^{\mathbf{\dagger }}$ can creates
one field states with  a generic wave functional $F$, as follows 
\begin{equation}
|n_{1}>=\int \mathcal{D}\phi \,F_{n}[\phi ]S^{\dagger }[\phi ]|0>
\label{stato1}
\end{equation}
with 
\begin{equation}
<n_{1}|n_{1}>=\int \mathcal{D}\phi \,F_{n}^{\ast }[\phi ]F_{n}[\phi ]=1
\end{equation}
The low index 1 in the label $n_{1}$,  means that the state (\ref{stato1}) is
an eigenstate of $\mathcal{N}$, with eigenvalue 1. It is easy to verify that 
$\mathcal{H}|n_{1}>=E_{n}\,\ |n_{1}>$ implies the Schr\"{o}dinger equation 
(\ref{eq:scroe}) (with $S$ replaced by $F$).  For certain choices of 
 the Hamiltonian $H$%
, the state with minimum energy could be of the type (\ref{stato1}), and 
we can  label it $|0_{1}>$. States with two
functional fields $|n_{2}>$  would lead to two identical (and decoupled)\
equations for  two different functionals $F_{1}[\phi _{1}]$ and $\
F_{2}[\phi _{2}]$: they would describe two parallel worlds that do not talk
to each other, as long as we restrict to the free\footnote{%
The Hamiltonian $H$  includes interaction in the common second quantization
sense, but it is free in the third quantization sense. This will become more
clear in the next section, when we will add a third quantization interaction.%
} Hamiltonian (\ref{hami}). .

A (Higgs) field that in second quantization is represented by a scalar field
operator $\hat{\phi}$  (see section 2), in the context of  a
third quantization can be  replaced by a proper combination of the functional
operators $S^{\dagger }[\phi ]$ and $S[\phi ]$. Namely,  
\begin{eqnarray}
\hat{\phi}(x) &\Rightarrow &\hat{\Phi}(x)\equiv \int \mathcal{D}\phi
\,S^{\dagger }[\phi ]\phi (x)S[\phi ]  \label{phi} \\
\hat{\pi}(x) &\Rightarrow &\hat{\Pi}(x)\equiv i\int \mathcal{D}\phi
\,S^{\dagger }[\phi ]\,\ \frac{\delta \,S[\phi ]}{\delta \phi (x)}.
\end{eqnarray}
One can verify that the commutation relations (\ref{comm1}) are satisfied
(using (\ref{anticom})). The operators $\hat{\Phi}$ and $\hat{\Pi}$ are
 common \
field operators (i.e. not functional operators), and until now the (\ref{phi})
can be considered an alternative\footnote{%
This means that  if  we restrict to a theory described by the Hamiltonian 
(\ref{hami}) (bilinear in the functional operators $S)$, and with 
 physical observables defined by $\Pi $ and $\Phi ,$ we have a theory that
is completely equivalent to a  quantum field theory. Departures from 
quantum field theory will
become manifest  when we will add to (\ref{hami}) an interaction with  the
insertion of several functional operators $S$.  }  representation of the
algebra (\ref{comm1}).

In the Heisenberg picture the functional field operator $S[\phi ]$ becomes
time dependent, and satisfies the quantum mechanical equation 
\begin{equation}
i\,\ \frac{d\,S[\phi ,t]}{d\,t}=\left[ \mathcal{H},S[\phi ,t]\right]
\end{equation}
from which we get 
\begin{eqnarray}
i\,\ \frac{d\,\hat{\Phi}(x,t)}{d\,t} &=&\left[ \mathcal{H},\hat{\Phi}(x,t)%
\right] =\hat{\Pi}(x,t)  \label{erenfest2} \\
\frac{d^{2}\,\hat{\Phi}(x,t)}{d\,t^{2}} &=&-i\,\ \frac{d\,\hat{\Pi}(x,t)}{%
d\,t}=-\left[ \mathcal{H},\hat{\Pi}(x,t)\right] =(\triangledown ^{2}-m^{2})%
\hat{\Phi}(x,t)  \label{erenfest3}
\end{eqnarray}

In particular we can find that the ordinary two point Green function of the
scalar field, applying the definition (\ref{phi}), 
\begin{equation}
G(x,t;y,t^{\prime })=<0_{1}|T\left\{ \Phi (x,t)\Phi (y,t^{\prime })\right\}
|0_{1}>
\end{equation}

where $T$ is the time ordered product. $|0_{1}>$   is the state with
minimum energy, with $\mathcal{N\,}|n_{1}>=|n_{1}>$,  and $F_{0}$ is equal
to (\ref{vacuum}), namely 
\begin{equation}
|0_{1}>=\int \mathcal{D}\phi \,F_{0}[\phi ]S^{\dagger }[\phi ]|0>
\label{vuoto}
\end{equation}
Then applying twice (\ref{erenfest2}), we can check the equation 
\begin{equation}
\left( \partial ^{2}+m^{2}\right) G(x,t;y,t^{\prime })=\delta
^{3}(x-y)\delta (t-t^{\prime })
\end{equation}

This confirms that the state $|0_{1}>$,in third quantization,  describes a
physical system that is equivalent to the one described by a scalar quantum
field theory with Green functions given by the rules of second quantization.
The concept of Green functions can be generalized to third quantization,
provided that we replace functions with functionals. For example, in
complete analogy with the second quantization, we can define a two-point
Green Functional 
\begin{equation}
G[\phi ,t;\,\phi ^{\prime },t^{\prime }]=<0|T\left\{ S[\phi ^{\prime
},t^{\prime }]S^{\dagger }[\phi ,t]\right\} |0>  \label{greenfunc}
\end{equation}
One can verify that it corresponds to the inverse functional operator
appearing in (\ref{fund}), once we define a proper time ordering prescription
to get rid of the pole singularities (the $i\,\varepsilon $ Feynman \
prescription).

\subsection{\protect\bigskip  The Hierarchy problem:
Interacting functional fields}

We have seen in the first section that the hierarchy puzzle could be solved
if we add a non linear term to equation (\ref{eq:scroe}).  One could
introduce it by hand  modifying (\ref{eq:scroe}), as in a non-linear Schr\"{o}%
dinger equation, and studying the phenomenological consequences. However
here we prefer to stick to quantum theory. In fact, we notice that adding a
non linear interaction reminds us the embedding of first quantization into 
second quantization, \textit{i.e.} the possibility of  interaction changing
the number of particles, that in the second quantization language
corresponds to operators involving more than two fields. In other words the
hierarchy puzzle can be seen as an hint to move forward with some  sort of
 third quantization beyond the second one.

Unfortunately, until now, we have not many clues to guess which type of
interaction to add. We have not yet any principle as powerful  as that of
gauge theories leading to unambiguous forms for the interactions.
Nevertheless, note that   (\ref{anticom}) implies that the mass dimension of
the functional field $S[\phi ]$ is not well defined since the functional
delta has the same dimensions of  the product of  an infinite number of
dirac delta. But we also know that 
\begin{equation}
\dint \mathcal{D}\phi \{iS^{\dagger }[\phi ,t],S[\phi ^{\prime },t]\}=\dint 
\mathcal{D}\phi \,\ \,\delta ^{\infty }[\phi -\phi ^{\prime }]=1,
\end{equation}
thus the dimension of the integral $\dint \mathcal{D}\phi $ is  the inverse
of a pair of $S.$  If we want to add a pair of functional fields  $%
S^{\dagger }S$ we are obliged to add an integral $\dint \mathcal{D}\phi $,
if we want to keep the action with the right mass dimensions. For instance $%
\int \mathcal{D}\phi S^{\dagger }[\phi ]S[\phi ]=\mathcal{N}$ is a
dimensionless operator. In the following, we restrict ourselves to
Hamiltonians that commutes with $\mathcal{N}$ , thus the number of fields is
conserved. The phenomenological consequences of additional insertions to the
Hamiltonian of  different powers of $ \mathcal{N}$ , $\mathcal{N}^{2}$
, etc.  are rather trivial. Instead, let us consider the operator\footnote{%
Even if not explicitly stated, we assume that the field $\phi $  can carry
internal indices of some internal symmetry. This justifies why we have
inserted the square $\phi ^{2}$ (to build an invariant operator), instead of 
$\phi $. Also $S$ can carry an index of an internal symmetry, but to
simplify the notation we remove this index.} $\mathcal{O}(x)=\int \mathcal{D}%
\phi \,S^{\dagger }[\phi ]\phi ^{2}(x)S[\phi ]$. $\mathcal{O}$  commutes
with $\mathcal{N}$ , it has well defined mass dimensions,  and  can be
used to build Hamiltonians with much more interesting  phenomenology. As
example, we will consider the Hamiltonian (\ref{hami})   with an additional
interaction of the form 
\begin{equation}
\Delta \mathcal{H=}\int d^{3}x\,\ \sum_{n}c_{n}\mathcal{O}(x)^{n}=\int
d^{3}x\,\ V(\mathcal{O}(x))  \label{inter}
\end{equation}
where $V$ is an arbitrary function, defined by the expansion above with
coefficients $c_{n}$. This interaction is local in the space-time
dimensions. It is understood that some renormalization prescription for the $%
c_{n}$ are  considered; for our purpose we can also look at the (\ref{inter})
as an effective hamiltonian, arising from a more fundamental theory
involving  different functional operators in addition to $S$.

\subsubsection{The two point Green functional in the non perturbative vacuum
of the Theory}

The insertion of (\ref{inter}), leads to a new action $\mathcal{A}$, given by

\begin{equation}
\mathcal{A=}\dint \mathcal{D}\phi \,dt\,\ S^{\dagger }[\phi ,t]\,\left( i%
\frac{\partial }{\partial t}+H\right) \,S[\phi ,t]+\,\int dt\,d^{3}xV\left(
\,\dint \mathcal{D}\phi ^{\prime }S^{\dagger }[\phi ^{\prime },t]\,\phi
^{\prime 2}(x)\,S[\phi ^{\prime },t]\right)  \label{azion}
\end{equation}

\bigskip

The Green functional (\ref{greenfunc}) at the  zeroth order approximation, \
as anticipated in the previous section, is given by 
\begin{equation}
G_{0}[\phi ,t;\,\phi ^{\prime },t^{\prime }]=i\left( i\frac{\partial }{%
\partial t}+H+i\,\varepsilon \right) ^{-1}[\phi ,t;\,\phi ^{\prime
},t^{\prime }]  \label{zeroorder}
\end{equation}
However the exact two point Green functional is affected by the interaction $%
V$ that can change the vacuum of the theory (in the following we have in
mind an analogous example in second quantization, the Nambu-Jona-Lasinio
model). Naively, taking the derivative $\delta \mathcal{A}/\delta S^{\dagger
}=0$, we  deduce the equation 
\begin{equation}
i\frac{\partial }{\partial t}S[\phi ,t]=\ H\,\ S[\phi ,t]+V^{\prime }\left(
\left\langle 0|\mathcal{O}|0\right\rangle \,\right) \int d^{3}x\text{ }\phi
^{2}(x)\,\,S[\phi ,t]  \label{meanfield}
\end{equation}
where, in the mean field approximation, we have replaced the interacting
term with its vacuum expectation value $<0|V^{\prime }(\mathcal{O})|0>\simeq
V^{\prime }(\left\langle \mathcal{O}\right\rangle )=V^{\prime }(v^{2})$. In
this approximation, we see that we still have a linear Schr\"{o}dinger
equation (\ref{meanfield}), as in  second quantization (\ref{eq:scroe}).
 The only  effect
of the interaction $V$, is to modify  the bare mass of
the field $\phi $. In fact, a mass $V^{\prime }(v^{2})$ adds directly to the
hamiltonian $H$, in (\ref{meanfield}), and thus modifies also the Green
Functional  (\ref{zeroorder}). Apart from 
this renormalization of the bare mass,
the physics described by the approximate equation (\ref{meanfield}), is 
identical to an ordinary quantum field theory. 

If we wish to
compute the exact Green functional, we have to compute all loop
contributions induced by the interaction $V$. These can be formally taken
into account in a non perturbative approach for composite operators (see
chapter 8 in \cite{miransky}): the exact Green functions arise from the
functional minimization of an effective action  $\Gamma \lbrack G]$, a
functional of the Green function $G$. $\Gamma $ is the sum of all (two-particle
irreducible)  vacuum loop diagrams. We will not repeat here the arguments
leading to the equations below\footnote{%
Also because the generalization of path integrals into an equivalent
mathematical object in third quantization is involved and unclear.}, instead
we assume a straightforward generalization of the discussion in \cite
{miransky}. We replace Green functions with Green functionals, and
integrations over space coordinates with functional integrations over the
function $\phi (x)$. As a result, the effective action for a composite
functional operator $S^{\dagger }[\phi ,t]\,S[\phi ^{\prime },t^{\prime }]$
 is a functional $\Gamma \lbrack G]$ of the Green functional 
(\ref{greenfunc})%
, that can be written\footnote{%
We assume $\left\langle 0\left| S(\phi )\right| 0\right\rangle =0.$}
(compare with eq. 8.47 of \cite{miransky}) 
\begin{equation}
\Gamma \lbrack G]=-iTr(\log (G^{-1}))-iTr(G_{0}^{-1}G)+\Gamma _{2}[G]
\label{eff}
\end{equation}
\bigskip where $\Gamma _{2}$ is given by all two-loop (and higher)
two-particle irreducible vacuum graphs.
\begin{figure}[h]
\center
\vskip  0.6cm
\epsfig{figure= 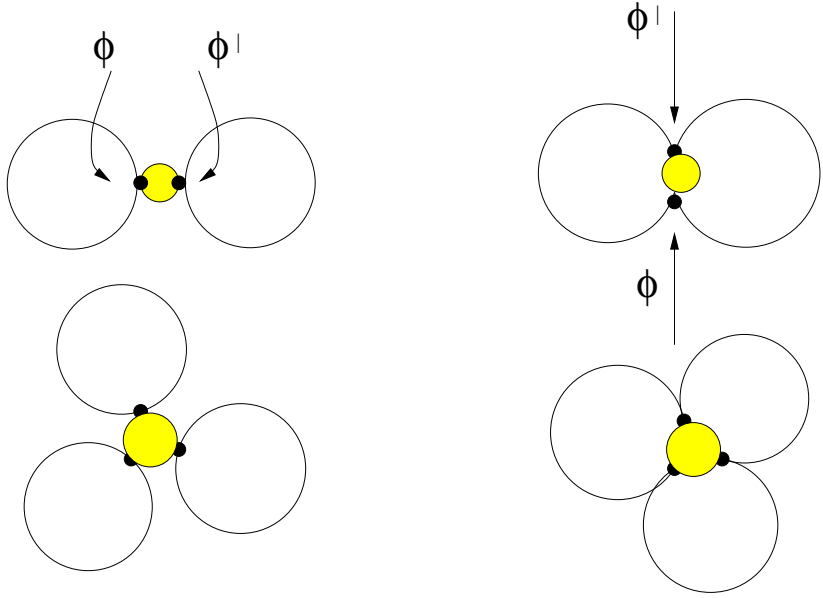, height= 8cm}
\caption{
Two particle irreducible feynman diagrams. The small black points stand
for distinct  points in the functional space $\protect\phi ,\protect\phi
^{\prime },...$. The lighter and bigger points stands for a point in the
ordinary space-time. Diagrams on the left, involve only Green functionals \
at the same functional point $\protect\phi $. i.e. $G[\protect\phi ,t;%
\protect\phi ,t]$. Instead on the right they only involve $G[\protect\phi ,t;%
\protect\phi ^{\prime },t]$ with $\protect\phi \neq \protect\phi ^{\prime }$%
.  }
\end{figure}
In the following (see Fig. 1),
we will  consider only vacuum loop diagrams, at first order in $V$, i.e.
with only one space-time point (represented in Figure 1, by the big grey
point).  $V$ contains several powers of the operator $%
\mathcal{O}(x)$; each $\mathcal{O}(x)$ contains a functional integration \
with respect a distinct function $\phi $. This explains why we have also
depicted few small dark points, each one representing a distinct functional
point $\phi ,\phi ^{\prime },\phi ^{\prime \prime },$ etc. For example, the
top diagram on the left  comes from the interaction 
term $\mathcal{O}^{2}$ . $\mathcal{O}^{2}$
 two functional integrations with respect the
two functions $\phi ,\phi ^{\prime }$ (depicted by the two distinct black
points) . In the same diagram there are two lines (i.e. two  Green
functionals or propagators), attached at the same black point. That is to
say each Green functional is evaluate at the same functional point 
($G[\phi,\phi ]$). 

 From the (\ref{eff}), and reminding the definition  (\ref{delta1})
of functional derivative, we can write the equation 
\begin{equation}
\frac{\delta \Gamma \lbrack G]}{\delta \,G}=iG^{-1}-iG_{0}^{-1}+\frac{\delta
\Gamma _{2}[G]}{\delta \,G}.  \label{equazione}
\end{equation}
This yields the exact functional $G$ that we want to compute.

We can distinguish two  type of contributions, when we take the derivative
of the functional, $\delta \Gamma _{2}/\delta G$. The first is obtained when
the derivative $\delta \,\ /\delta G[\phi ,\phi ^{\prime }]$ acts on the
Green functional $\,G[\phi ^{\prime \prime },t;\,\phi ^{\prime \prime },t]$,
computed at the same function $\phi ^{\prime \prime }$, (and same time $t$).
For example, those coming from the diagrams on the left of Figure 1. After
the integration over $\phi ^{\prime \prime }$, these yield a term
proportional to a functional Dirac delta $\delta ^{\infty }[\phi -\phi
^{\prime }]$. A second one, arises when the derivative acts on a Green
functional $G[\phi ^{\prime \prime \prime },t;\,\phi ^{\prime \prime },t]$
with different functions $\phi ^{\prime \prime \prime }$ and $\phi ^{\prime
\prime }$ (but same time $t$): the full derivative takes the form 
\begin{equation}
\begin{array}{c}
\frac{\delta \Gamma _{2}[G]}{\delta \,G[\phi ,t;\phi ^{\prime },t^{\prime }]}%
=A(v^{2})\,\,\,\,\,\,\delta (t-t^{\prime })\int d^{3}x\,\phi ^{2}[x]\,\delta
^{\infty }(\phi -\phi ^{\prime })+ \\ 
\\ 
+B(v^{2})\,\,\,\ \ \delta (t-t^{\prime })\int d^{3}x\,\phi ^{2}[x]\,\,G[\phi
,t;\,\phi ^{\prime },t]\phi ^{\prime 2}[x]+...
\end{array}
\end{equation}
Where the dots $...$ stand for all other possible combinations of the Green
functionals $G$ arising from the same interaction $V$. For brevity, we do
not list them all, since they are unnecessary  for the discussion below.
The functions $A(v^{2})$ and $B(v^{2})$   have appeared as a result of all
integrations of the functionals $G$ over $\ \phi $. Their
explicit dependance on the vacuum expectation value $v=<0|\hat{\phi}(x)|0>$,
 can be computed  only if a specific interaction $V$  is chosen, and
after the evaluation of all functional integrals. Among these, we note for
instance  
\begin{equation}
\dint \mathcal{D}\phi \,G[\phi ,t;\phi ,t]\,\,\,\phi \lbrack x]^{2}\equiv
\left\langle 0\left| \hat{\phi}^{2}(x)\right| 0\right\rangle =v^{2}
\label{vev}
\end{equation}
In this step we have assumed\footnote{%
The Green function is obtained as the inverse of the kinetic  operator of
the classical action (\ref{zeroorder}). The choice (\ref{bound})%
corresponds to a particular $i\varepsilon $ prescription for the pole
singularities in (\ref{zeroorder}). These prescriptions set the boundary
conditions for $G$ at $t^{\prime }=t$.} the boundary condition at $t^{\prime
}=t$ 
\begin{equation}
G[\phi ,t;\phi ^{\prime },t]=\,\ F_{0}(\phi )F_{0}^{\ast }(\phi ^{\prime }).
\label{bound}
\end{equation}
Where $F_{0}$ is of the form (\ref{vacuum}). With this prescription,  we get a
vacuum with one (and only one) higgs field (see eq. (\ref{vuoto}) and eq. (\ref
{greenfunc})).  As in ordinary quantum field theory, where  the evaluation of
the composite operator $\hat{\phi}^{2}(x)$ needs a suitable renormalization
prescription, we understand that suitable  counterterms have been added to
ensure a finite result in the integration (\ref{vev}).  Note that $v$  is the
true vev of the theory\footnote{%
We remind that we look for space-time translation invariant solutions, such
that $\,G[\phi ,t;\,\phi ^{\prime },t^{\prime }]\rightarrow G[\phi ,t;\,\phi
^{\prime },t^{\prime }]$ when $\phi (x)\rightarrow \phi
(x+x_{0}),\,t\rightarrow t+t_{0},\,t^{\prime }\rightarrow t^{\prime }+t_{0}$
and $\phi ^{\prime }(x)\rightarrow \phi ^{\prime }(x+x_{0})$ then the
integral $\dint \mathcal{D}\phi \,\ $\ in eq.(\ref{vev}) yields a constant $v$
independent of $x.$}; i.e. all radiative corrections  are taken into
account, since $G$ is  the full Green Functional of the theory$.$ The full
equation (\ref{equazione}) for the Green functional $G[\phi ,t;\,\phi ^{\prime
},t^{\prime }]$ can be written 
\begin{equation}
\begin{array}{c}
G[\phi ,t;\,\phi ^{\prime },t^{\prime }]^{-1}=i\frac{\partial }{\partial t}%
+H+i\,\varepsilon +A[v^{2}]\,\,\,\,\,\,\delta (t-t^{\prime })\int
d^{3}x\,\phi ^{2}[x]\,\delta ^{\infty }(\phi -\phi ^{\prime })+ \\ 
\\ 
+B\left[ v^{2}\right] \,\,\,\ \ \delta (t-t^{\prime })\int d^{3}x\,\phi
^{2}[x]\,\,G[\phi ,t;\,\phi ^{\prime },t]\phi ^{\prime 2}[x]+...
\end{array}
\label{miransky}
\end{equation}
We can distinguish two parts in the right-hand side: the one proportional to 
$A[v^{2}]\,\ \delta ^{\infty }(\phi -\phi ^{\prime })$ is similar to a mass
term in $H$ (see eq. (\ref{eq:scroe})) and can be reabsorbed into a
redefinition of the bare mass of the field $m_{0}^{2}\phi ^{2}\rightarrow
m_{1}^{2}\,\phi ^{2}=(m_{0}^{2}+A[v^{2}])\,\phi ^{2}$.  One can recognize
that in the mean field approximation (eq. (\ref{meanfield})) $%
A[v^{2}]=V^{\prime }[v^{2}]$  (and $B=0$).  The last term is not
proportional to $\delta ^{\infty }(\phi -\phi ^{\prime })$; for the moment,
as a first approximation, we assume $B[v^{2}]$ to be very small and
negligeable.  The equations  (\ref{miransky}) and (\ref{vev}) are a system of
coupled equations, from which we would like to determine $v$ and $G$. Let us
forget for a moment (\ref{vev}).  We look for solutions $G$ of (\ref{miransky}),
treating  $v$ as a free parameter. We recognize that (\ref{miransky}), (with $%
B=0$) is the same equation that we would get in absence of the interaction $V
$ (see the section before), apart from a redefinition of the bare mass%
\footnote{%
We stress that this redefinition occurs at the level of the bare mass,
before any integration over loop momenta and before the action of the
renormalization group equations.} 
\begin{equation}
m_{0}^{2}\phi ^{2}\rightarrow m_{1}^{2}\,\phi ^{2}=(m_{0}^{2}+
A[v^{2}])\,\phi ^{2}.  \label{bare}
\end{equation}
This implies that we can compute $G$ following the more familiar  procedure
of second quantization . We remember that $G$ is defined in (\ref{greenfunc}).
In particular we can compute the vacuum (\ref{vev}) of this modified
hamiltonian as follows. We first calculate the full effective potential of
an ordinary quantum field theory with Hamiltonian $H$, but   with modified
bare masses (\ref{bare}).   The minimization of this non perturbative 
effective potential would give 
\begin{equation}
v^{2}=\mathcal{V}^{2}\left( m_{1}^{2},g_{1}^{2},...\right) =\mu ^{2}\left(
m_{1}^{2},g_{1}^{2},...\right) /\lambda \left( m_{1}^{2},g_{1}^{2}\right)
+...  \label{minimo}
\end{equation}
$\ v\ $ comes from a function $\mathcal{V}$ of  the bare parameters $%
m_{1}^{2},g_{1}^{2},...$  at the large scale $\Lambda $. The function $%
\mathcal{V}$   can be computed in the case of a perturbative quantum field
theory; exploiting the renormalization group equations (RGE) one usually 
finds $v$ as a simple function of the common Higgs potential parameters $%
\mu $ and $\lambda $. In any case if $ m_{1}$ and the other mass parameters
at the high energy scale are of the order of the large scale $\Lambda $ also
 $v$ \ from eq. (\ref{minimo}) is of order $\Lambda $. This is the origin of
the Hierarchy problem: if any of the mass parameters of the theory entering
the $\mathcal{V}$ is large, then also $v$ is large. Only with a fine tuning
of those large masses $m_{1}$ etc. one can get a small vev $v.$ Now let us
see how this paradox can be solved in the context of third quantization. In
this case $m_{1}$ , the bare mass at the scale $\Lambda $, is not a free
parameter but it is itself a function of the vev $ v$  through equation 
(\ref{vev}) and (\ref{bare}). Using these equations the (\ref{minimo}) becomes 
\begin{equation}
v^{2}=\mathcal{V}^{2}\left( (m_{0}^{2}+A[v^{2}]),g_{1}^{2},...\right) 
\label{minimo2}
\end{equation}
It is understood that the above equation is exact, i.e. it includes all
quantum corrections: the function $\mathcal{V}$  includes the exact (all
loops) dependance from the parameters $m_{1}^{2},g_{1}^{2},...$ given at
the very high energy. It is clear that the equation (\ref{minimo2}) can have
solutions with $v<<\Lambda $: even if the function $\mathcal{V}^{2}$ is
rather simple (e.g. polynomial).  Mathematical examples could be easily
built if   $A$  contains logarithms or exponentials of the vev $v$. 
The hierarchy  arises exclusively from the interactions at the third
quantization  level, that in the (\ref{minimo2}) are represented by the
function $A$. This interplay between second and third quantization is
parallel to a similar connection between first and second quantization. An 
interesting 
example, is provided by  the hydrogen
atom. In this specific physical system,  one finds the minimum of an
Hamiltonian (in first quantization), with the fine structure constant 
 $\alpha $, that can be calculated
 in second quantization. Namely,  firstly we solve  the Schr\"{o}%
dinger equation (first quantization), and we find the electron wave function
$\psi$.
$\psi$ contains the  characteristic scale $p$ of the virtual momentum
exchanged between the proton and the electron. This $p$ is clearly a
function of the input parameter $\alpha $. But this input parameter $\alpha $
is in turn a function of $p$, since the fine structure constant  is 
obtained (in second quantization) from the $\beta $ function and the  RGE,
evaluated    at the characteristic scale $p.$ The interplay between first
and second quantization can be summarized as follows: $p$ is obtained from $%
\alpha $ applying the common procedure of first quantization. In turn 
$ \alpha $ is a function of the characteristic scale $p$, and can be computed
using second quantization and the RGE equations. This connection leads to a
non linear equation that is absolutely equivalent to (\ref{minimo2}): in fact 
second quantization (and the minimization of the full effective potential)\
is used to derive the  characteristic scale $v$  as a function $\mathcal{V}
$ of the bare parameters $m_{1}^{2},g_{1}^{2},$  while third quantization
(through the interaction $V$) is necessary  to calculate $m_{1}^{2}$  (the
bare mass) as a  function of $v$. This leads to (\ref{minimo2}).

 Even if the phenomenological consequences of a third quantization need an
 appropriate study  of well defined and realistic models, we can
anticipate some generic possible consequences implied by this scenario. These
could lead to new and exotic  experimental signatures. One of these can be
inferred by the additional terms proportional to $B$, that appeared in 
(\ref{miransky}) (and that we have neglected until now). If we consider non
trivial vacua, then  the functional $G$ solving (\ref{miransky}) could  be
different from that one of an ordinary quantum field theory. Let us consider
the modification to the solution $G$, induced by the additional term $B$. If 
$G_{0}$ indicates the Green functional at $B=0$, then eq.(\ref{miransky}) \
gives us the first order  perturbative  correction

\begin{equation}
G_{1}^{-1}[\phi ,t;\phi ^{\prime },t^{\prime }]=G_{0}^{-1}[\phi ,t;\phi
^{\prime },t^{\prime }]+B(v^{2})\ \delta (t-t^{\prime })\int d^{3}x\phi
^{2}(x)\phi ^{\prime 2}(x)G_{0}[\phi ,t;\phi ^{\prime },t].  \label{greenuno}
\end{equation}
Plugging the (\ref{bound}) in (\ref{greenuno}), we get the Green functional $%
G_{1}$ describing the time evolution of the quantum state (the wave
functional) from the initial time $t^{\prime }$ to the final time $t$. To
the ordinary Hamiltonian $H$ , we have to add an operator of the form  
\begin{equation}
\Delta H=B(v^{2})\int d^{3}x\left( \,\ \hat{\phi}^{2}(x)|0><0|\hat{\phi}%
^{2}(x)\right)   \label{newint}
\end{equation}
The insertion of the vacuum projector $|0><0|$ is rather unusual . However 
note, that if $G_{0}^{-1}$ is the renormalizable hamiltonian of the Standard
Model, it  includes the Standard Model symmetries, like  the baryon and the
lepton number. Thus the vacuum $\ F_{0}$ \ (\ref{vacuum}), (obtained from $%
G_{0}^{-1}$, eq. (\ref{zeroorder})) is invariant under   these symmetries. Thus  also the new
interaction (\ref{newint}) looks symmetric. All accidental symmetries of the
renormalizable standard Model are preserved. At first glance, we can also
anticipate few possible experimental signatures. Let us write down the Green
function involved in the scattering of two Higgses into two Higgses $h\,\
h\rightarrow h\,\ h$. For simplicity we consider the case where $H$  in (\ref
{zeroorder}) contains only the kinetic term,  thus the only interaction in
this process, comes from $\Delta H$ (\ref{newint}), and it appears only at the
first order in $B(v^{2})$. The four point Green function is  
\begin{equation}
g(x_{1},x_{2},x_{3},x_{4})=<0|\phi (x_{1},t_{1})\phi (x_{2},t_{2})\phi
(x_{3},t_{3})\phi (x_{4},t_{4})|0>
\end{equation}
with the $t_{1,2}\rightarrow +\infty $ in the far future and $%
t_{3,4}\rightarrow -\infty $ in the far past. At the first non trivial order
we get 
\begin{eqnarray}
& & g(x_{1},x_{2},x_{3},x_{4}) = \\
   &=&\int_{-\infty }^{+\infty }dt<0|\phi
(x_{1},t_{1})\phi (x_{2},t_{2})\Delta H(t)\phi (x_{3},t_{3})\phi
(x_{4},t_{4})|0>= \\
&=&B(v^{2})\int_{-\infty }^{+\infty }d^{3}x\,dt<0|\phi (x_{1},t_{1})\phi
(x_{2},t_{2})\hat{\phi}^{2}(x,t)|0><0|\hat{\phi}^{2}(x,t)\phi
(x_{3},t_{3})\phi (x_{4},t_{4})|0>
\end{eqnarray}
where $\ \Delta H$ is inserted in the middle, given that $%
t_{1,2}>>t>>t_{3,4},$ and we have to respect the time ordering
prescription. $ h \, h \rightarrow h \, h $ 
is mediated  by $\Delta H$  as if it were an
ordinary $\hat{\phi}^{4}$ operator. On the other hand, the same effective $%
\hat{\phi}^{4}$ interaction, added to the original interaction, would also
modify a process like $h\rightarrow h\,\ h\,\ h$; but this process is not
affected by $\Delta H$, as it can be easily checked, because 
 we know that $<0|%
\hat{\phi}^{2}(x,t)\phi (x_{4},t_{4})|0>=0$. In other words, in such a
scenario, an anomalous self-interaction of the Higgs boson could be proven
by the existence of  two discrepant precision measurements\footnote{%
This discrepancy would inevitably prove the failure of ordinary quantum
field theory in favor of  an extension with a new type (third quantized)
interactions.} of the same effective coupling constant but  in two
different physical processes.

\section{Conclusions}

The unexpected experimental success of the renormalizable Standard Model
puts severe constraints to any attempt to solve the Hierarchy problem by
adding new particles in the low energy spectrum of the theory. Their real
(or virtual) production is ruled out, at least in the region around the
electroweak scale. We have seen that the main difficulty is due to the
theoretical assumption that the time evolution operator of quantum states
with very different vacuum expectation values of $\phi $ are described by
the same linear operator, the Hamiltonian H. While the Standard Model
Hamiltonian, seems to be very accurate to describe particle physics
phenomenology (in experimental  tests concerning  only quantum states very
close to the electroweak vacuum), it seems to be unable to estimate the
energy of vacua with very different expectation values of the field operator 
$\phi $. This paradox becomes more affordable if we accept the possibility
that the time evolution of the quantum states is described by a non-linear
Schr\"{o}dinger equation. Motivated by this example, we have explored how
such a modification can be achieved still in the context of quantum
mechanics, but with an embedding of second quantization into some sort of
third quantization. We already know that first quantization arises as an
approximation of second quantization: namely, if we neglect the interaction
between particles and we consider only two-point Green functions. In fact,
two-point Green functions satisfy the Schr\"{o}dinger equation of first
quantization. Similarly, we can imagine that second quantization (i.e.
quantum field theory) arises as an approximation of  third quantization,
where the role of points and fields is now played, respectively, by fields
and functional fields. As for first quantization, second quantization is
obtained in the limit when only the two-point Green functional is relevant.
In this limit, the Hamiltonian of quantum field theory is simply the inverse
of the full two-point Green functional of the third quantization theory.
Thus, to derive the Hamiltonian of second quantization from the full third
quantization theory, we have used the formalism of the effective action of
composite operators. We have found that the interaction of third
quantization $A$, can significantly change the equation (\ref{minimo2})
without adding any new particles in the low energy spectrum. This minimal
impact in the low energy spectrum  is, to our opinion, in better agreement
with the experimental evidence. Even if we have not yet shown that the idea
suggested in this paper can lead  to a realistic and well defined theory,
we have   anticipated few possible experimental signature,  from a simple and
naive analysis of our mathematical example above. Namely,  a new type of
operators could appear in the effective Hamiltonian of second quantization.
Beside the common self interaction of the operators $\phi ^{4}(x)$, one
could also have some exotic operators of the form $\phi ^{2}(x)|0><0|\phi
^{2}(x)$ with the insertion of the vacuum projector. This interaction would
lead to some anomalous (and non- linear) interactions in scattering
processes involving the Higgs particle.

\end{document}